# Experimental Evaluation of Passive Optical Network Based Data Centre Architecture

Azza E. A. Eltraify, Mohamed O. I. Musa, Ahmed Al-Quzweeni and Jaafar M.H. Elmirghani

*School of Electronic and Electrical Engineering, University of Leeds, Leeds, LS2 9JT, UK*
E-mail: {scaeae, m.musa, ml13anaq, j.m.h.elmirghani}@leeds.ac.uk

**ABSTRACT**
Passive optical networks (PON) technology is increasingly becoming an attractive solution in modern data centres as it provides energy efficient, high capacity, low cost, scalable and flexible connectivity. In this paper we report the implementation of a PON based data centre architecture that provides high resilience and high speed interconnections by providing alternative communication routes between servers in different racks. Each rack is divided into several groups of servers and connects to other racks and the Optical Line Terminal (OLT) through a set of server that acts as relay servers. We implement the switching and routing functionalities within servers using 4x10GE Xilinx NetFPGA, and demonstrate end-to-end communication using IP cameras live video streaming over up to 100 km optical connections through WDM nodes and the PON network.

## 1. INTRODUCTION

The increasing demand for higher data rates and power efficiency in data centres have been the focus of research and development in data centres in recent years. Several studies have been carried out in optimising energy efficiency and architectures in data centres and core networks connecting them [1] - [10]. However, there is a need to transform these theoretical advancements into practical and tangible systems.

Our previously designed PON based data centre network architecture shown in Figure 1 [11] proposed a server centric architecture that facilitates high speed interconnections among servers by dividing the whole data centre into a set of PON cells that contain an aggregation of several servers, ONUs and an OLT. Inside each cell, servers are divided into racks and within a rack, servers are grouped in multiple groups. Servers of different groups in the same rack are connected via an optical backplane [12]. One of the servers' groups is connected to an OLT and is responsible for the rack's communication with the OLT, while other groups are connected to other racks in the cell. Inter-rack communication is established by using one of the group's servers via a direct connection with the rack of the destination server. Connections between the OLT and racks can be provisioned either using a coupler (TDM) or via an AWGR (WDM) to provide more bandwidth [11], [13] - [15].

One of the major problems in data centres nowadays is keeping up with the increasing demand for higher data rates, faster processing and scalability. Several applications have been implemented to mitigate these problems and help fulfil the massive demand in data rates, one of the solutions was the introduction of NetFPGAs, a Field Programmable Gate Arrays (FPGA) based networking solution. FPGAs were introduced to accelerate the switching, routing and processing of network data. The NetFPGA Platform is an open source hardware and software platform which contains a large Xilinx FPGA that is programmed with user-defined logic and implements the PCI interface to a host processor. The platform includes Static RAMs (SRAMs) that operate synchronously with the FPGA as well as Double Data Rate (DDR2) SDRAM devices that operate asynchronously with the FPGA. A quad-port physical-layer transceiver (PHY) is provided enabling the platform to send and receive packets. The NetFPGA platform now has three main operating boards that are supported by the NetFPGA community which are NetFPGA-SUME, NetFPGA-1G-CML and the NetFPGA-10G [16]. In this paper the NetFPGA-10G is used as a Network Interface Card (NIC) for the servers. It is an FPGA-based PCI Express board with four 10-Gigabit SFP+ interfaces, an x8 gen1 PCIe adapter card incorporating Xilinx's Virtex-5 TX240TFPGA. It is considered to be ideal for high bandwidth applications [16].

This work aims to experimentally demonstrate the implementation of the proposed data centre architecture and furthermore demonstrate end-to-end communication of different data centres over long distances through various platforms. The end-to-end connectivity is shown using IP cameras live video streaming over up to 100 km optical connections through WDM nodes and the PON network.

## 2. ARCHITECTURE IMPLEMENTATION

To implement the architecture shown in Figure 1, three racks were employed. Each set has three servers grouped together representing a different rack/ data centre. The communication between the servers within the same rack uses a 10 Gbps Cisco Switch and the communication between racks is set up using an optical link where media converters are used to convert the electrical signal to optical. The communication between servers and the flow of data between different racks is illustrated in Figure 2. Figure 3 shows the experimental implementation of the proposed architecture. As an initial implementation, we employ an electronic switch instead of the optical polymer backplane and use the operating system of the servers to forward packets.

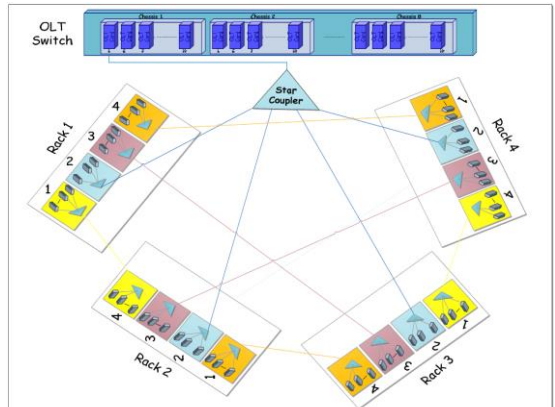

*Figure 1. Server Centric PON based data centre network architecture [17]*

Servers were set up using two different operating systems. Each operating system serves a different purpose. The majority of the servers operated under the open source Debian based Ubuntu Operating system commonly used in networking. Servers acting as gateways operated using MikroTik OS; a specialized Linux Kernel operating system that provides routing functionalities in order to oversee packet routing from and to racks.

The IP network is organized into three networks based on the number of racks and communication links between these racks. The connections between servers utilise various paths through different racks to ensure reliability. Hence, in order to realise the data connectivity between servers in different racks, the packet routing has been configured in servers by the use of MikroTik. Figure 2 illustrates the networks and IP assignment where each rack has a server that oversees packet routing from and to the rack configured with two IP addresses, one for intra-rack communication and another for inter-rack communication. Other servers in the rack have been assigned a gateway IP address for inter-rack communication.

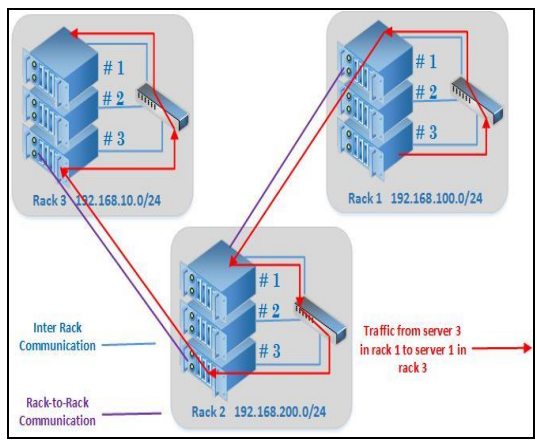

*Figure 2. Intra-rack and inter-rack communication in the proposed PON based data centre architecture*

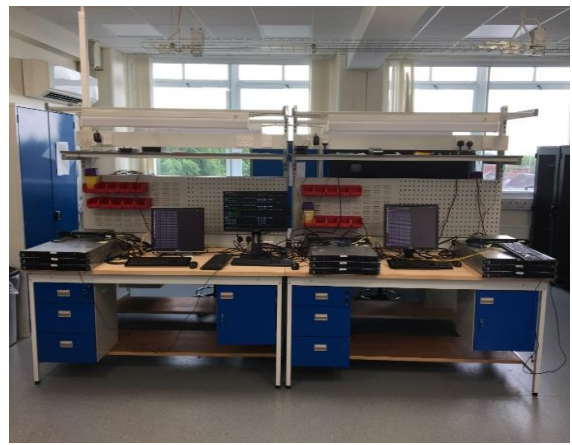

*Figure 3. Server Centric Architecture Experimental Implementation*

To facilitate communication between multiple data centres located in multiple cities, an IP over DWDM core network, spanning 100 km over optical fibre links, was implemented as shown in Figure 4. Each core node is an MRV/ADVA DWDM node with up to 100 Gbps capabilities. Multiplexers/Demultiplexers are used on the C-band supporting 80 wavelengths, and EDFAs and dispersion compensators are used to improve the quality of the optical signal over long distances. This setup allowed testing data transmission and delay over various long distances representing actual distances between cities in a typical core network. [18].

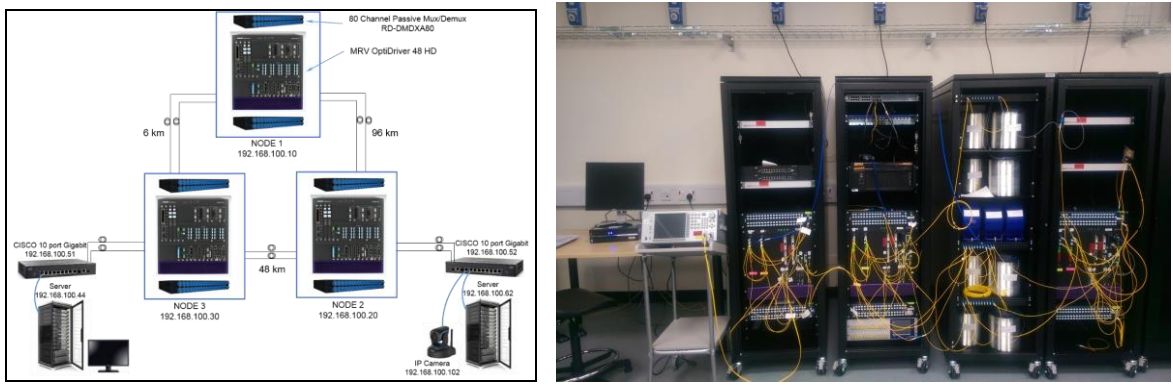

*Figure 4. IP/WDM Core Network Nodes*

In order to demonstrate end-to-end communication, all previous set ups were connected together optically as shown in Figure 5, where the IPTV Camera is connected to the proposed architecture and streams live video through the Fog data centre architecture, over the IP/WDM Core networks nodes, and finally to the OLT/ONU where a display is connected to show the video stream.

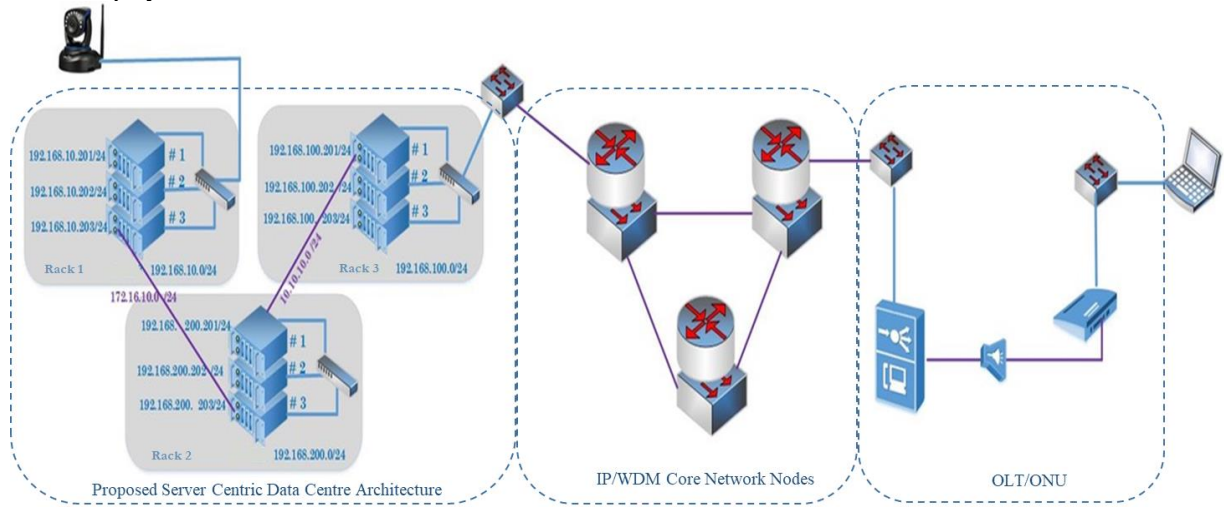

*Figure 5. The implemented end-to-end System of the Proposed Server Centric Data Centre Architecture, IP/WDM Core Network Nodes and OLT/ONU*

## 3. RESULTS

In order to achieve the full architecture setup, three components were implemented: (a) Servers group setup (b) IP/WDM Core Network Nodes, and (c) OLT/ONU. The end-to end system connectivity was demonstrated using real time video streaming over a distance of over 100 km. The video was streamed using two different platforms, an IoT camera and another IPTV camera where both resulted in real time high quality live video stream.

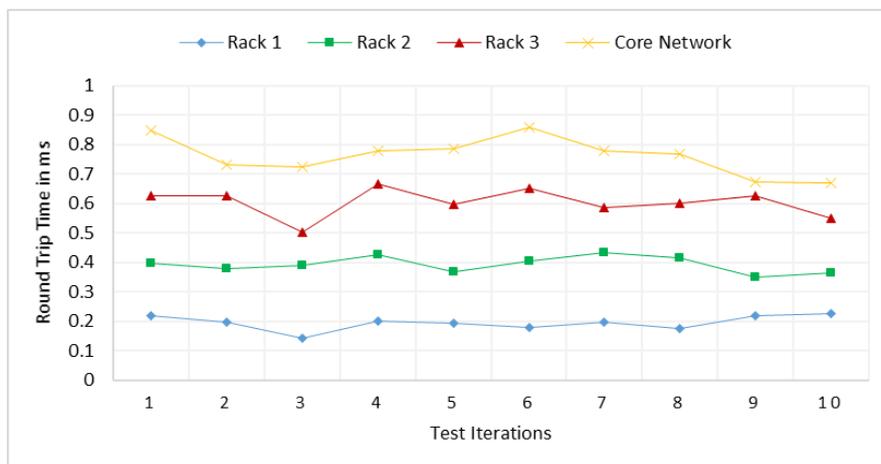

*Figure 6. The Round Trip time of all nodes in the system for 10 iterations.*

Network connections and packet routing were tested using the ICMP protocol and route tracing. ICMP signals were sent from each server to all servers and were successfully received with 0% packet loss. Traceroute was sent from the server where the camera was connected in the Fog data centre to the end node in the core network, in order to measure the latency in each hop. The test was performed by sending packets and measuring the round trip time (RTT) of each hop. This signal was sent 10 times, each iteration resulting in different results for the RTT of 150 results in total. Figure 6 shows the 10 tests iterations and the results of each hop's average RTT.

The system demonstrated low latency, with RTT less than 1 ms (ranging from 0.144 ms to 0.857ms) as shown in Figure 6, which is significantly low considering the long distance and the difference in platforms.

## 4. CONCLUSIONS

This paper has described an experimental evaluation of a proposed Service Centric PON based data centre architecture that is cost and power efficient. The performance was demonstrated through the use of ICMP signals and a video stream over the server centric architecture, IP/WDM Nodes and OLT/ONU communication. The use of NetFPGA boards for NIC communication provided high speed and high data rate optical interconnections between servers. Future implementation includes adding an Arrayed waveguide grating router (AWGR) to provide wavelength routing between servers [14], [15] leading to a further reduction in cost and power consumption.

ACKNOWLEDGEMENTS

The authors would like to acknowledge funding from the Engineering and Physical Sciences Research Council (EPSRC), INTERNET (EP/H040536/1) and STAR (EP/K016873/1). All data are provided in full in the results section of this paper.